\documentclass[aps,preprint,showkeys,footnote,10pt]{revtex4}
\usepackage{amsfonts}
\usepackage{amsmath}
\usepackage{amssymb}
\usepackage{graphicx}
\usepackage{subfigure}
\usepackage{txfonts}
\usepackage{makeidx}

\newtheorem{theorem}{Theorem}
\newtheorem{acknowledgement}[theorem]{Acknowledgment}

\begin{document}

\title{\textbf{Electron-phonon dynamics in 2D carbon based-hybrids XC (X=Si,
Ge, Sn)}}
\author{L. B. Drissi$^{1,2,*}$, N. B-J. Kanga$^{1},$ S. Lounis$^{3}$ , F.
Djeffal$^{4}$, S. Haddad$^{5}$}
\affiliation{{\small 1-LPHE-Modeling \& Simulations, Faculty of Science, Mohammed V
University in Rabat, Rabat, Morocco}}
\affiliation{{\small 2-CPM, Centre of Physics and Mathematics, Faculty of Science,
Mohammed V University in Rabat, Rabat, Morocco}}
\affiliation{{\small 3-Peter Gr\"{u}nberg Institut and Institute for Advanced Simulation,
Forschungszentrum J\"{u}lich and JARA, J\"{u}lich, Germany}}
\affiliation{{\small 4-LEA, Department of Electronics, University Mostefa
Benboulaid-Batna 2, Batna, Algeria}}
\affiliation{
{\small 5-}{\small LPMC, Faculty of Science of Tunis University Tunis El
Manar, Tunis, Tunisia}}
\affiliation{{\small *} {\small email:  ldrissi@fsr.ac.ma}}

\begin{abstract}
The effect of the presence of electron-phonon (e-ph) coupling in the $SiC$, $%
GeC$ and $SnC$ hybrids is studied in the framework of the ab initio
perturbation theory. The electronic bang gap thermal dependence reveals a
normal monotonic decrease in the $SiC$ and $GeC$ semiconductors, whereas $%
SnC $ exhibits an anomalous behavior. The electron line widths were
evaluated and the contributions of acoustic and optical phonon modes to the
imaginary part of the self-energy were determined. It has been found that
the e-ph scattering rates are globally controlled by the out-of-plane
acoustic transverse mode $ZA$ in SiC while both $ZA$ and $ZO$ are overriding
in $GeC$. In $SnC$, the out-of-plane transverse optical mode $ZO$ is the
most dominant. The relaxation lifetime of the photo-excited electrons shows
that the thermalization of the hot carrier occurs at $90fs$, $100fs$ and $%
120fs$ in $SiC$, $GeC$ and $SnC$ respectively. The present study properly
describes the subpicosecond time scale after sunlight illumination using an
approach that requires no empirical data. The results make the investigated
structures suitable for providing low cost and high-performance optical
communication and monitoring applications using 2D materials.
\end{abstract}

\maketitle

\section{Introduction}

The graphene synthesis revealed exceptional properties for this material
such as, high ambient temperature mobility, ambipolar effect, Klein
tunneling, anomalous quantum Hall effect, etc \cite{24}. This Breakthrough
suggested many opportunities for the creation of other two dimensional (2D)
materials. Among them, stanene, germanene and silicene are of particular
interest because of their compatibility with current silicon technology \cite%
{sii}. Similar to graphene, the electrons in these 2D materials behave as
massless Dirac fermions and exhibit a linear dispersion-band near the Fermi
level due to their $sp^{2}$ electronic configuration \cite{75}. The
successful growth of Si, Sn and Ge- 2D monolayers has led to extensive
investigation into the search for new hexagonal materials, such as
phosphorene, borophene, aluminene and ZnO, with great potential for
applications.

The materials consisting of column IV-elements of the periodic table are
gapless in the absence of spin orbit coupling. This fact makes them less
interesting for optoelectronic applications such as photovoltaic cells and
light emitting diodes, which require a material with a nonzero band gap \cite%
{18}. To overcome this deficiency, columns IV-IV hybrid materials have been
introduced \cite{topsakal,grr}. The combination of silicon and carbon atoms
creates silicene/graphene monolayer (SiC). This hybrid has a 2D planar
honeycomb structure rather than a buckled one due to the strong $\pi $-bond
through its perpendicular p$_{\mathbf{z}}$ orbitals\textbf{\ }\cite{19}-\cite%
{111}. On the other hand, germanene/graphene hybrid (GeC) films are another
column IV binary compound which can be prepared using deposition techniques
such as laser ablation \cite{211} and radio frequency reactive sputtering in
Ar/CH$_{4}$ \cite{212}. This material possesses excellent electro-catalytic
properties and is promising for fuel cell and lithium--oxygen battery
applications \cite{GeCaa}. Both GeC and SiC sheets are direct band gap
semiconductors and show strong excitonic effects with high binding energy
\cite{1,2}. Meanwhile, the SnC monolayer, resulting from the combination of
graphene and stanene, exhibits an indirect band gap. All these 2D
graphene-based materials are good candidates for optoelectronic applications
and are very promising for nanotechnology applications requiring catalytic
performance for reduction and oxidation.

In a large group of semiconductors, a monotonic decrease of the energy gap
has been observed as the temperature increases \cite%
{Eliashyambo10,Eliashyambo11}. This well understood phenomenon does not
occur in some exceptional materials where band gaps exhibit two kinds of
anomalous temperature dependence. In the non-monotonic anomalous case, the
gap first increases at low temperatures and then decreases at high
temperature \cite{Eliashyambo12}. In the monotonic anomalous case, the gap
increases continuously with temperature. This behavior is observed for some
perovskites \cite{Eliashyambo14}, copper halides \cite{Eliashyambo15} and
lead chalcogenides \cite{Eliashyambo16,Eliashyambo17}. Despite the abundance
of experimental evidence, the understanding of anomalous gap dependence is
explained by different, and sometimes even contradictory arguments, such as%
\textbf{\ }the role played by harmonic and anharmonic contributions \cite%
{Eliashyambo17,abc}.\textbf{\ }

The temperature dependence of the electronic energies originates partially
from the volume expansion \cite{bardeen} and results mainly from the
electron-phonon (e-ph) interactions at constant volume \cite{ttte}.\textbf{\
}In the latter case, the major contribution of the effect of e-ph
interaction proves to be the most difficult to compute from first principles
due to the huge number of k-points in the Brillouin zone required for
convergence \cite{reee}\textbf{. }This can be avoided by using schemes based
on Wannier functions. The spatial localization of the electronic and lattice
Wannier functions allows to compute only a limited set of electronic and
vibrational states as well as the corresponding e-ph interaction matrix
elements from first principles. This requires low costs of a standard phonon
dispersion calculation for a large number of $k-$points in the Brillouin
zone \cite{barr}.

Attention to the study of properties related to electron-phonon couplings
(EPC)\textbf{\ }using ab-initio computation has become more and more common
for 2D materials. The obtained results subsequently explain many phenomena
observed in\textbf{\ }honeycomb materials\textbf{. }The presence of Kohn
anomalies in graphene and in the low buckled silicene is associated with
their highest-branched optical modes \cite{grapp, eph1}. However, a
negligible EPC is observed in germanene which shows very small square of the
EPC matrix-element \cite{eph1}. In stanene, the electron-phonon process is
mainly dominated by the out-of-plane optical mode ($ZO$) leading to a hot
carrier thermalization occurring at 250 fs \cite{nouss}. In silicene, the
low energy of out-of-plane acoustic phonon mode $(ZA)$, derived from the
weak sp$^{2}$ bond between its Si atoms, contributes to the high scattering
rate and significant degradation in electron transport \cite{Mos2}. With
access to the first-principles EPC, the effect of the scattering rate on the
performance of MoS$_{2}$ and WSe$_{2}$ FET for a channel length of 10 nm
shows a ballistic of $83\%$ and $75\%$ for WSe$_{2}$\ and MoS$_{2}$ devices
respectively \cite{MWs}.

Therefore, understanding the EPC from imaginary self energy, electron life
time and carrier mobility in 2D hybrids is highly desirable and essential in
order to facilitate their further applications. Considering the interest for
graphene-based hybrids: GeC, SiC and SnC, this work studies the
electron-phonon coupling involved in the temperature effect on the gap
behavior. Using an ab initio approach based on the density functional theory
(DFT), we investigate the behavior of the electronic bang gap thermal
dependence in the three carbon based hybrids in the presence of
electron-phonon interaction at constant volume. With the exception of SnC,
which shows a non-monotonic anomaly, both GeC and SiC exhibit no anomalous
behavior in their gap variation. To calculate the electron lifetime in these
2D materials, the imaginary part of electron's self energy around the Fermi
level and the corresponding scattering rate are determined in detail. The
phonons contribution to the linewidth for an electron initially at the
conduction band minimum (CBM) and the valence band maximum (VBM), as well as
the variation of their corresponding lifetime with respect to the electron
energy are discussed.\textbf{\ },

The present paper is organized as follows. In section II, we present the
computational method required for the calculation of the reported data. In
section III, we start by a comparative description of the temperature effect
on the gap value. Then, we report a complete description of the behavior of
SiC, GeC and SnC electron linewidth with respect to the electron energy and
the temperature. From the plot of the scattering rate, we compute the
electron lifetime and elucidate the role of phonon modes. Finally, in
section IV, we end with a conclusion.

\section{Computational method}

\begin{figure}[tbph]
\begin{center}
\includegraphics[scale=0.35]{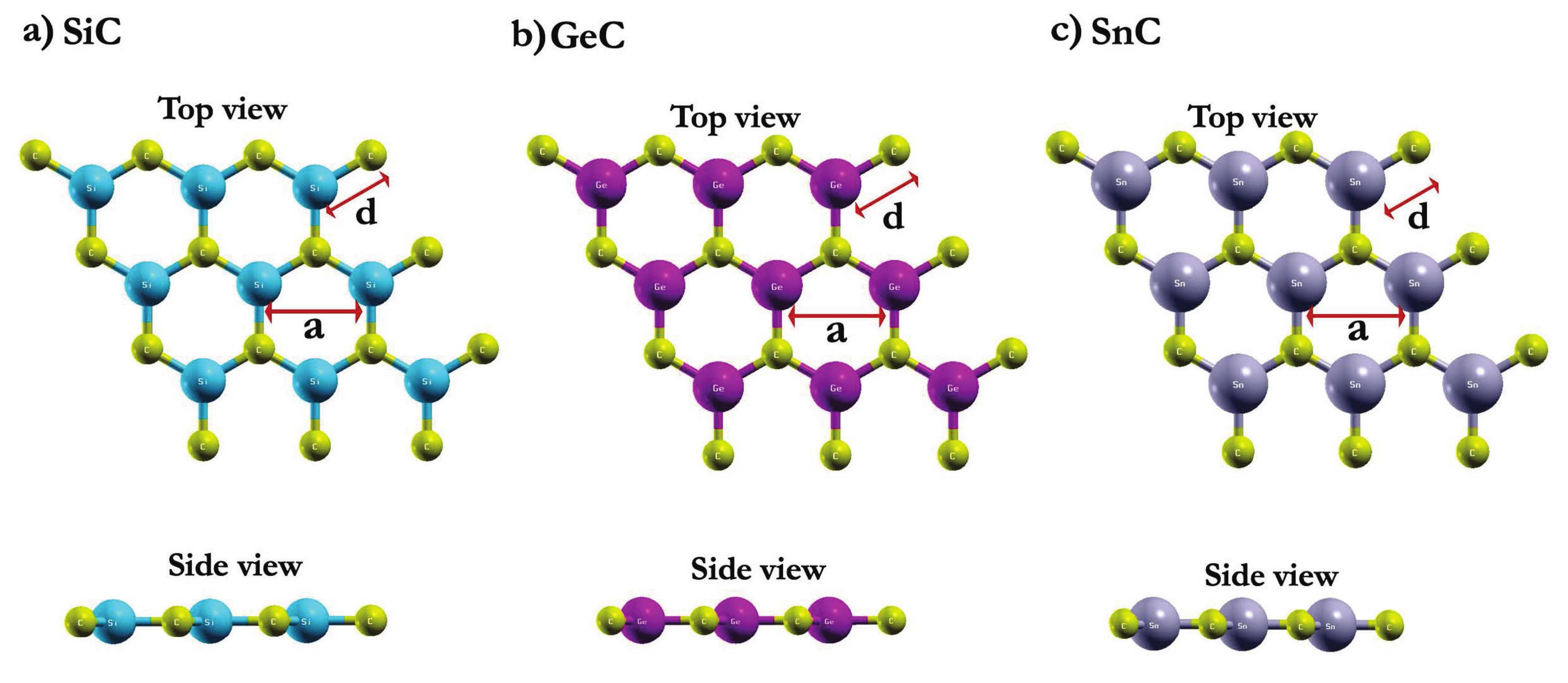}
\end{center}
\caption{{\protect \small Top and side view of atomic structure of planar
hybrids (a) SiC with the lattice parameter a=3.09 \AA \ and the bond length
d=1.785 \AA , (b) GeC with a=3.245 \AA \ and d=1.965 \AA , (c) SnC with
a=3.58 \AA \ and d=2.258 \AA .}}
\label{fig1}
\end{figure}

In this work, three planar carbon-based hybrids are studied. The unit cell
of each crystal contains 2 atoms: C and X (X=Si, Ge and Sn). As displayed in
Fig.\ref{fig1} , each carbon atom is bonded to three X-atoms forming
hexagonal structures.

Employing norm-conserving pseudo-potentials, the generalized gradient
approximation of Perdew-Zunger exchange-correlation functional is used
considering a plane-wave basis set with a kinetic-energy cutoff of 60 Ry
\cite{gga}. The electronic states are computed using a k-sampling grid of $%
32\times 32\times 1$ in the Brillouin zone based on the Monkhorst-Pack
scheme. The subsequent computational recipe is based on the scheme proposed
in \cite{ttte}.

The vibrational frequencies, $\omega _{q\lambda }$, and the derivatives of
the self-consistent Kohn-Sham potential with respect to the atomic
displacements are calculated on $16\times 16\times 1$ q-point grids using
density functional perturbation theory (DFPT) \cite{barr} as implemented in
the Quantum-Espresso code \cite{manu30}. These parameters are needed to
evaluate the e-ph coupling matrix elements.

The many body perturbation theory $(MBPT)$ is used to describe the behavior
of electronic states as a function of temperature \cite{mbpt}. The
electron-phonon coupling is perturbatively treated \cite{yambo47}. We
consider the Fan and Debye-Waller self-energy terms corresponding
respectively to the first and second order of the Taylor expansion in the
displacement of the nuclei. The Green function is written as \cite{ttte}:

\begin{equation}
G_{n\mathbf{k}}(\omega ,T)=\left[ \omega -\epsilon _{n\mathbf{k}}-\Sigma_{n%
\mathbf{k}}^{Fan}(\omega ,T)-\Sigma_{n\mathbf{k}}^{DW}(T)\right] ^{-1}
\label{green}
\end{equation}%
where $\epsilon _{n}$ is the Kohn-Sham ground-state eigenenergies of the
frozen atoms, $\sum^{Fan}$ is the Fan contribution to the self energy given
by:

\begin{equation}
\Sigma_{n\mathbf{k}}^{Fan}=\sum_{n^{\prime },\mathbf{q}\lambda }\dfrac{%
|g_{nn^{\prime }\mathbf{k}}^{\mathbf{q}\lambda }|^{2}}{N}\left[ \dfrac{N_{%
\mathbf{q}\lambda }(T)+1-f_{n^{\prime} \mathbf{k-q}}}{i\omega -\epsilon
_{n^{\prime }\mathbf{k-q}}-\omega _{\mathbf{q}\lambda }}+\dfrac{N_{\mathbf{q}%
\lambda }(T)+f_{n^{\prime }\mathbf{k-q}}}{i\omega -\epsilon _{n^{\prime }%
\mathbf{k-q}}+\omega _{\mathbf{q}\lambda }}\right],  \label{fan}
\end{equation}%
and $\Sigma^{DW}$ is the Debye-Waller term expressed as follows \cite{loiui}:

\begin{equation}
\Sigma _{n\mathbf{k}}^{DW}=-\dfrac{1}{2}\sum_{n^{\prime }\mathbf{q}\lambda }%
\dfrac{\Lambda _{nn^{\prime }\mathbf{k}}^{\mathbf{q}\lambda }}{N}\left[
\dfrac{2N_{\mathbf{q}\lambda }(T)+1}{\epsilon _{n\mathbf{k}}-\epsilon
_{n^{\prime }\mathbf{k}}}\right] ,  \label{dw}
\end{equation}%
with $N_{\mathbf{q}\lambda }$ and $f_{n^{\prime }}$ represent the
distribution functions of Bose-Einstein and Fermi-Dirac, while $\omega
_{\lambda }$ and $N$ are the phonon frequencies and\ the number of $\mathbf{q%
}-$points taken randomly to better map out the phonon transferred momentum
\cite{yambo47}. $\Lambda _{nn^{\prime }\mathbf{k}}^{\mathbf{q}\lambda }$ is
the second order electron-phonon matrix elements.

The first order electron-phonon matrix elements $g_{nn^{\prime }\mathbf{k}}^{%
\mathbf{q}\lambda }$, which represent the probability amplitude for an
electron to be scattered with emission or absorption of phonons are given
by:
\begin{equation*}
g_{nn^{\prime }\mathbf{k}}^{\mathbf{q}\lambda }=\langle \psi _{n^{\prime },%
\mathbf{k-q}}|\partial _{\mathbf{q}\lambda }V|\psi _{n,\mathbf{k}}\rangle ,
\end{equation*}%
where $n^{\prime }$ and $n$ are initial and final electron band indices with
wavevectors $\mathbf{k}$ and $\mathbf{k-q}$, respectively, and $\partial _{%
\mathbf{q}\lambda }$ is the derivative of the self-consistent potential
associated with a phonon of the wavevector $\mathbf{q}$ in the branch $%
\lambda $. To calculate the imaginary part of the electron's self-energy
using the Electron-Phonon coupling using Wannier functions code (EPW) \cite%
{epw}, the matrix elements were evaluated over a grid of $360\times \
360\times \ 1$ $\mathbf{q}$ points.

Assuming the quasiparticle approximation $(QPA),$ one can expand in
first-order the self-energy frequency dependence around the bare energies.
It follows that the temperature dependent quasi-particle energy is given by
\cite{ttte}:

\begin{equation}
E_{n\mathbf{k}}(T)\approx \epsilon _{n\mathbf{k}}+Z_{n\mathbf{k}}(T)\left[
\Sigma_{n\mathbf{k}}^{Fan}(\epsilon _{n\mathbf{k}},T)+\Sigma_{n\mathbf{k}%
}^{DW}(T)\right]  \label{energy}
\end{equation}%
where the renormalization factor is expressed as follows:

\begin{equation}
Z_{n\mathbf{k}}(T)=\left[ -\left. \dfrac{\partial R\sum_{n\mathbf{k}%
}^{Fan}(\omega )}{\partial \omega }\right \vert _{\omega =\epsilon _{n%
\mathbf{k}}}\right] ^{-1}.
\end{equation}

We note that the imaginary part of the Fan's self-energy ($\sum_{n\mathbf{k}%
}^{Fan}$) provides information on the rate of e-ph processes responsible for
hot carrier thermalization, which is related to the intrinsic quasiparticule
lifetime. The real part, however, informs us about the energy
renormalization shift. Thus, in order to obtain the associated e-ph
scattering rate, one should first compute the linewidth ($\Gamma _{n\mathbf{k%
}})$ \cite{epw}:
\begin{equation}
\tau _{n\mathbf{k}}^{el-ph}=\dfrac{2\Gamma _{n\mathbf{k}}}{\hslash },
\label{SR}
\end{equation}%
which just means that the quasiparticle lifetime corresponds to the inverse
of the scattering rate $\tau ^{-1}$.

Eq.\ref{energy} states that the single particle energy $E_{n\mathbf{k}}$
depends on temperature as follows \cite{ttte}:
\begin{equation}
E_{n\mathbf{k}}(T)=\epsilon _{n\mathbf{k}}+\Delta E_{n\mathbf{k}}(T)
\label{ener}
\end{equation}%
The energy $\Delta E_{n\mathbf{k}}(T)$ involves two contributions, namely:
\begin{equation}
\Delta E_{n\mathbf{k}}(T)\approx \left. \Delta E_{n\mathbf{k}}(T)\right
\vert _{har}+\left. \Delta E_{n\mathbf{k}}(T)\right \vert _{anhar}.
\label{contri}
\end{equation}%
The first term results from pure electron-phonon interactions and represents
the harmonic contribution at constant volume. The second term corresponds to
the effects flowing from the lattice expansion (variable volume). According
to equation \ref{contri}, it is obvious that the harmonic and anharmonic
terms have some contribution in the energy shifting. At constant volume, the
effect of the electron-phonon interaction is usually the major contribution,
so by setting $\left. \Delta E_{n\mathbf{k}}(T)\right \vert _{anhar}<<\left.
\Delta E_{n\mathbf{k}}(T)\right \vert _{har},$ only the harmonic effect will
be considered.\newline
In order to understand the gap behavior as function of temperature, one
should first compute the generalized Eliashberg spectral function. The
spectral function $g^{2}F_{n\mathbf{k}}(\omega )$ is expressed as follows
\cite{ttte}:
\begin{equation}
g^{2}F_{n\mathbf{k}}(\omega )=\dfrac{1}{N}\sum_{n^{\prime }\mathbf{q}\lambda
}\left[ \frac{|g_{nn^{\prime }\mathbf{k}}^{\mathbf{q}\lambda }|^{2}}{%
\epsilon _{n\mathbf{k}}-\epsilon _{n^{\prime }\mathbf{k-q}}}-\dfrac{1}{2}%
\dfrac{\Lambda _{nn^{\prime }\mathbf{k}}^{\mathbf{q}\lambda }}{\epsilon _{n%
\mathbf{k}}-\epsilon _{n^{\prime }\mathbf{k-q}}}\right] \delta (\omega
-\omega _{\mathbf{q}\lambda })  \label{eliashberg}
\end{equation}%
$\Lambda _{nn^{\prime }\mathbf{k}}^{\mathbf{q}\lambda }$ gives information
about the phonon mode contributions to the electron-phonon harmonic term $%
\left. \Delta E_{n}(T)\right \vert _{har}$ given by:
\begin{equation}
\left. \Delta E_{n}(T)\right \vert _{har}=\int d\omega g^{2}F_{n\mathbf{k}%
}(\omega )\left[ N_{\mathbf{q}\lambda }(T)+\dfrac{1}{2}\right] .
\label{deltaE}
\end{equation}
Now that the correction to the single particle state $E_{n\mathbf{k}}$ is
related to the Eliashberg function, one can estimate the correction to the
band gap, $E_{g}(T)$, by defining band-edge Eliashberg function, $F_{g}$,
which is simply defined \cite{ttte}:
\begin{equation*}
F_{g}=F_{CBM}-F_{VBM},
\end{equation*}%
where the sub-indices $CBM$ and $VBM$ refer to a given conduction and
valence state. The equation \ref{deltaE} reveals that a positive (negative)
region of $g^{2}F$ would then lead to an increasing (decreasing) gap as
function of temperature in contrast to a negative region stands for a
decreasing $E_{g}(T)$.

\section{Results and discussion}

\subsection{Temperature effect on the band gap}

\begin{figure}[tbph]
\begin{center}
\includegraphics[scale=0.450]{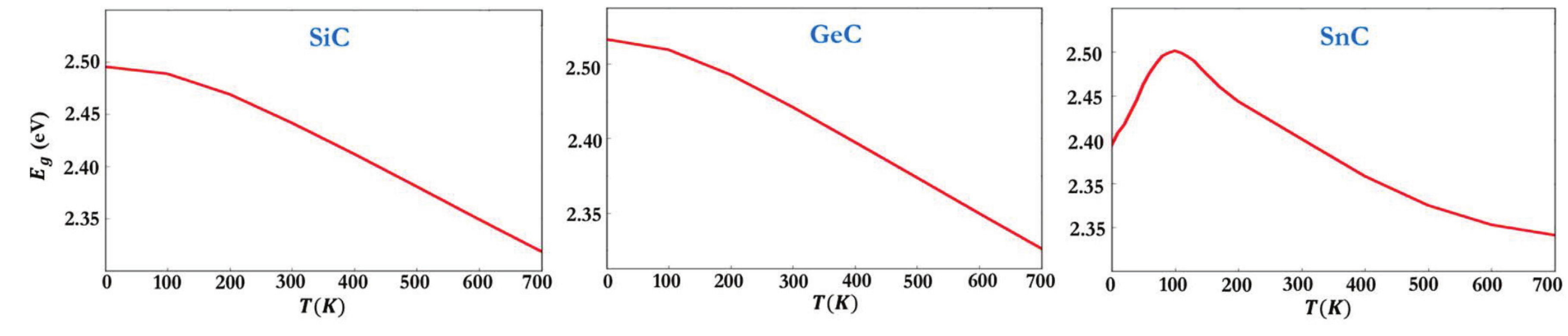}
\end{center}
\caption{{\protect \small Band gap measurement as a function of the
temperature for the three hybrids SiC, GeC and SnC.}}
\label{gap1}
\end{figure}

This part investigates the electron-phonon interaction in order to calculate
and describe the temperature dependence of the electronic gap $(E_{g})$ for
SiC, GeC and SnC hybrids. Fig.(\ref{gap1}) displays the variation of the
band gap with respect to the temperature ($T$) for the three carbon-based
hybrids.\newline
The value of the bandgap of both SiC and GeC show a monotonic decrease as a
function of temperature. It follows that these two materials are
semiconductors with normal behavior. Meanwhile, SnC shows a different
behavior. Indeed, by varying the temperature from $0K$ to $100K$ , the band
gap of the SnC configuration also increases as plotted in Fig.(\ref{gap1}).
Thus, SnC has an anomalous trend in this temperature range. However, for a
temperature above $100K$, the gap of SnC starts to decrease. This gap
behaviour makes SnC hybrid a semi-conductor of non-monotonic anomaly kind.

\begin{figure}[tbph]
\begin{center}
\includegraphics[scale=0.40]{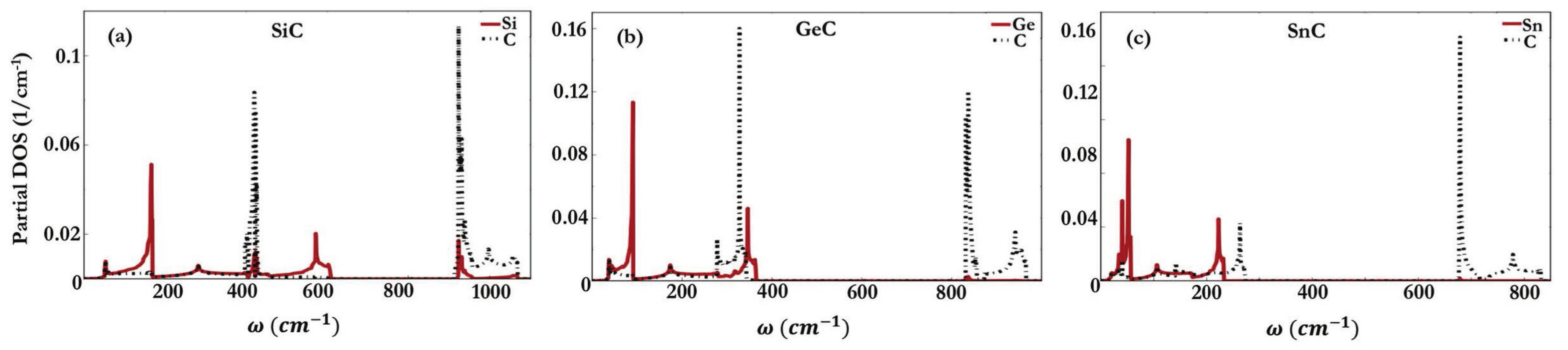}
\end{center}
\caption{{\protect \small Projected phonon density in terms of phonon
frequency for VBM (blue dashed line), CBM (red dotted line) and band edge
(black solid line) corresponding to SiC, GeC and SnC sheets.}}
\label{eliash}
\end{figure}
The projected phonon density of state in Fig.(\ref{eliash}) reveals that the
heavier is the atom the smaller is his contribution to the optical modes. In
the three hybrids, all optical modes are mainly dominated by carbon atoms,
while the acoustical modes are mostly dominated by the heaviest atoms,
namely Si, Ge and Sn in SiC, GeC and SnC respectively.

\begin{figure}[tbph]
\begin{center}
\includegraphics[scale=0.40]{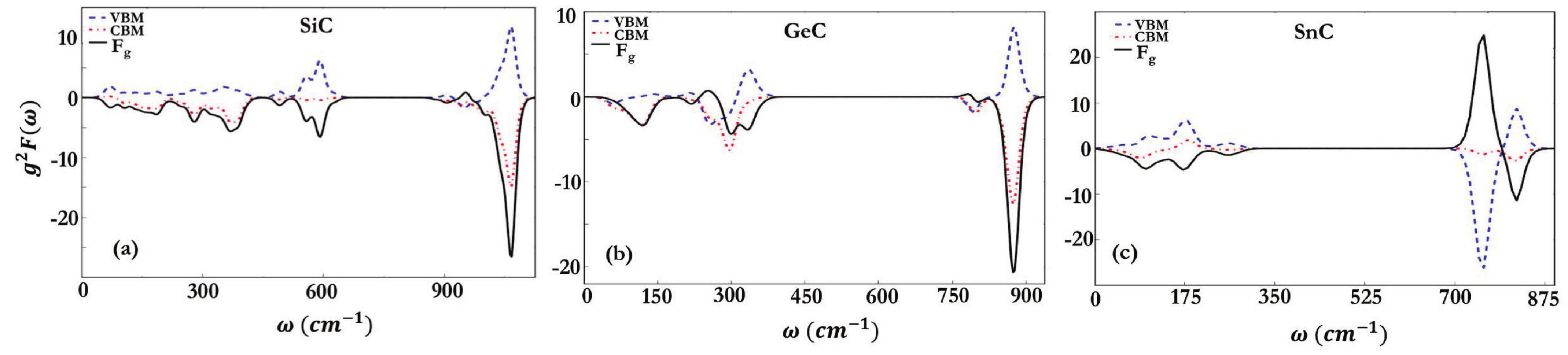}
\end{center}
\caption{{\protect \small Electron-phonon Eliashberg spectral function }$%
g^{2}F_{n\mathbf{k}}${\protect \small \ as a function of phonon frequency for
VBM (blue dashed line), CBM (red dotted line) and band edge (black solid
line) corresponding to the three hybrids.}}
\label{elly}
\end{figure}

To shed more light on the gap behavior for the three binary compounds, Fig.(%
\ref{elly}) plots the generalized Eliashberg spectral function\textbf{\ }for
the CBM and VBM states to visualize the e-ph coupling strength for a given
state $|n\mathbf{k}\rangle $ in SiC, GeC and SnC hybrids. Both SiC and GeC
show a conventional semiconductor behavior for all phonon frequencies in
agreement with the normal plot of their $E_{g}(T)$ previously highlighted in
Fig.(\ref{gap1}). When increasing the temperature, the band gap of SiC and
GeC exhibits the same behavior like the bulk Si and Ge \cite{cardona}.
However, the Eliashberg function describing the SnC hybrid has two relevant
regions. More precisely, the function behaves as in conventional
semiconductor for all acoustic modes and higher frequency optical modes,
while it shows an anomalous trend for lower frequency optical modes. The
pace of the Eliashberg function explains the anomalous non-monotonic
behavior of the SnC band gap as a function of temperature.

Generally, the Eliashberg function, defined in eq(\ref{eliashberg}) is
positive for the valence band maximum and negative for conduction band
minimum. This opposite trend leads to the reduction of the gap. Regarding
equation \ref{deltaE}, a positive region of $g^{2}F$ corresponds to a
temperature increasing of the gap energy\textbf{\ }while its negative region
stands for a decreasing of $E_{g}(T)$. For the carbon based hybrids, one
deduces that in SiC, the VBM and CBM Eliashberg functions are all normal for
all phonon modes leading to a gap decreasing. For GeC, the CBM contributions
are negative for all phonons while for the VBM, the contribution for the
Eliashberg function are positive except for phonon with frequency around $%
250cm^{-1}$. Therefore, the resulting Eliashberg function arising from VBM
and CBM contribution show a gap decreasing. Finally, the curve associated to
the SnC sheet shows a strong anomalous behavior for the VBM due to the
contribution of the high optical phonon modes. This anomalous behavior leads
to the non monotonic variation of the gap when increasing the temperature.

\subsection{Electron linewidth vs temperature}

\begin{figure}[tbph]
\begin{center}
\includegraphics[scale=0.3]{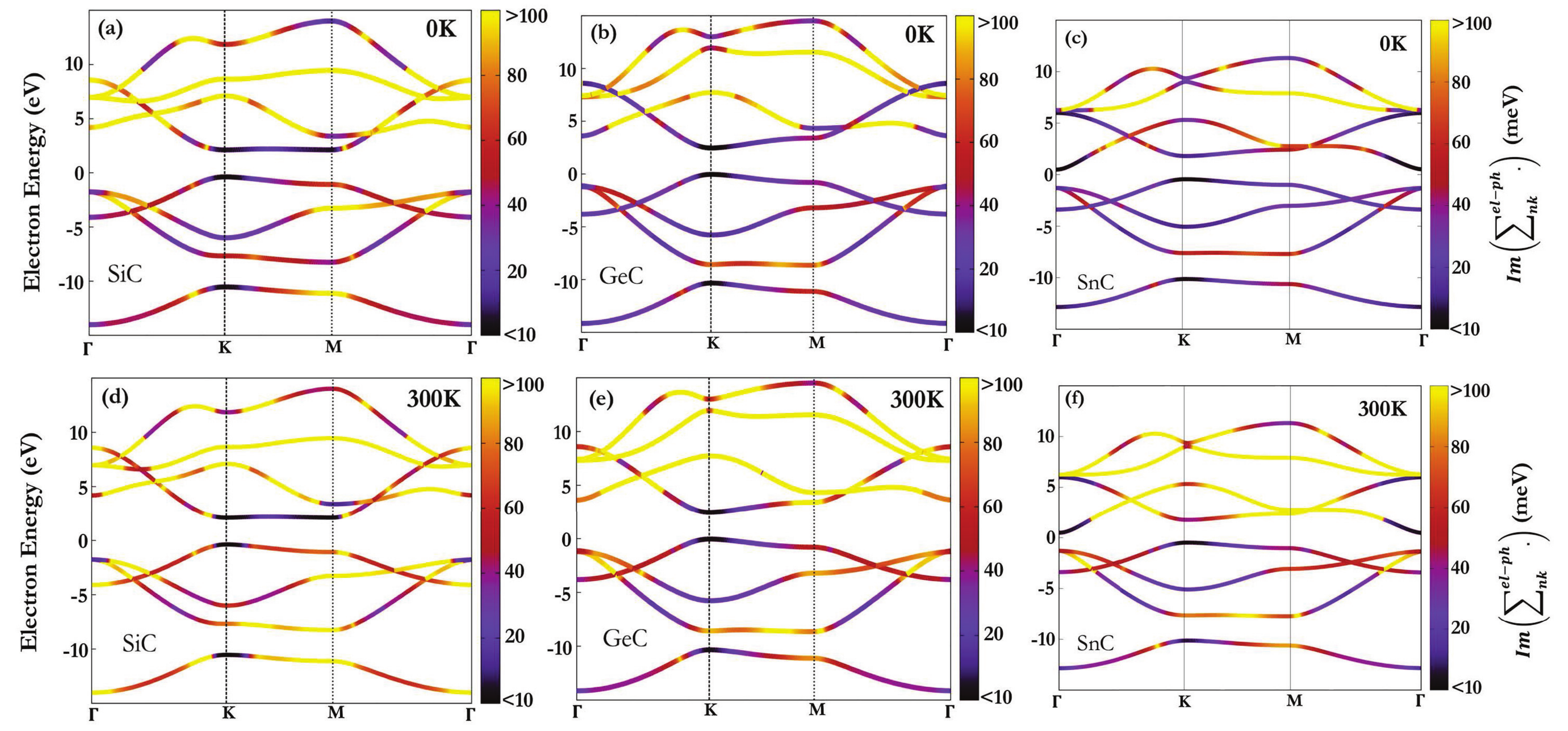}
\end{center}
\caption{{\protect \small Band structure of the hybrids SiC, GeC and SnC,
together with a color map of }$(\mathrm{Im}(\sum_{n\mathbf{k}}^{e-ph})).$%
{\protect \small at 0K (in the first line) and 300K (in the second line).}}
\label{line1}
\end{figure}

The following part investigates the electron linewidth arising from the
electron-phonon interaction and determines the electron lifetime for all
structures. Fig.(\ref{line1}) displays the pooling of the electronic band
structure of stanene and the imaginary part of the self energy $\Gamma _{n%
\mathbf{k}}$ on a fine electron wavevector k-grid for two different
temperatures namely, $0K$ and $300K$. The temperature $300K$ is chosen
rather than other temperatures in order to investigate hot carrier behaviour
at room temperature.

At $0K$, the variation of electron linewidth as a function of the electron
energy is plotted in Fig.(\ref{line1}) along k-points on the high symmetry
line for SiC, GeC and SnC respectively. The band structures of SiC, GeC and
SnC show that the states in SnC are more closer from each other compare to
SiC and GeC explaining the smaller linewidth for SnC with respect to SiC and
GeC, which is more observable in the unoccupied states.

Recall that the imaginary part $\Gamma _{n\mathbf{k}}$ of the Fan's
self-energy provides information on the rate of e-ph processes responsible
for hot carrier thermalization \cite{gius}. The following focus on the
carriers thermalization and their lifetime around the Fermi level for the
three monolayers.

At $300K$, Figs.(\ref{line1}-d,-e,-f) show that the imaginary part of the
self energy is more sensitive\textbf{\ }to the temperature in the conduction
bands compared to valence bands. This is due to the fact that, the carrier
in the conduction band are free to move which makes them very susceptible to
any external perturbation such as temperature. This sensitivity also depends
on the gap value as it gets increased when the gap value is decreasing.
Indeed, when increasing the temperature, the electron linewidth in the SiC
structure changes less significantly with respect to its two counterparts
GeC and SnC having smaller gaps. Moreover, compared to GeC, the linewidth in
SnC is more sensitive to temperature at the vicinity of the $M$-point.
However, far of this symmetric point and around the Fermi level, the thermal
sensitivity is quite similar in both GeC and SnC. The very small density of
state makes the imaginary self energy very small and unchangeable in term of
temperature dependence compared to other energy levels. The obtained results
are in good accordance with the high thermal sensitivity observed for bulk
silicon and bulk germanium due to their small gap values of $1.17eV$ and $%
0.74eV$ respectively \cite{sige}. Finally, the used method gives accurate
variation of the linewidth with respect to the electron energy for the three
carbon based hybrids as this method allows to extract information about the
linewidth for each energy level. This is promising for determining the exact
relaxation time and engineering hot carriers cells \cite{hot}.

\subsection{Scattering rate and lifetime}

\begin{figure}[tbph]
\begin{center}
\includegraphics[scale=0.35]{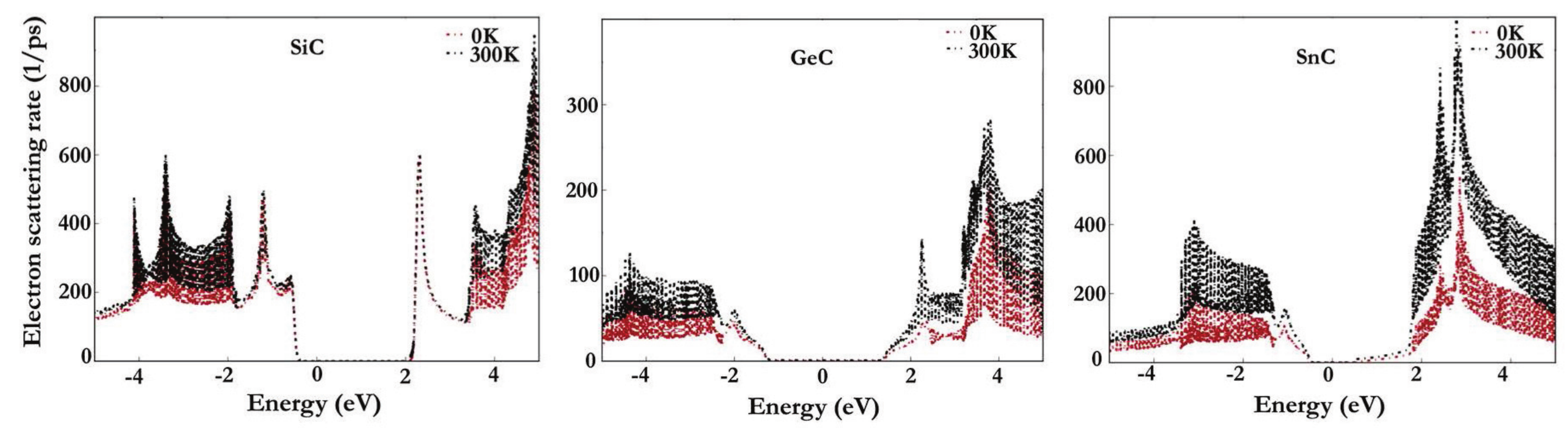}
\end{center}
\caption{{\protect \small The scattering rate associated with the (e-ph)
self-energy corresponding to SiC, GeC and SnC hybrids at the temperatures 0
and 300 K. }}
\label{x004}
\end{figure}

Based on eq(\ref{SR}), the electron-phonon scattering rate is plotted with
respect to the electron energy in Fig.(\ref{x004}) for the three hybrids
SiC, GeC and SnC. Only the electron-phonon contribution is considered as it
is the most dominant near the Fermi level compared to the electron-electron
interaction. Indeed, as reported in \cite{eph111,eph2}, hot electrons
essentially interact with phonon because of the small density of state and
the low energy range around the Fermi level.

Fig.(\ref{x004}) shows a very small change of the scattering rate in term of
temperature for the SiC hybrid$.$ The shifting is more observable in GeC
compared to SiC, and the displacement of the scattering rate is much more
visible for SnC. All these observations are in good agreement with the
results obtained for bulk germanium, bulk silicon \cite{sige} and diamond
\cite{diam}.
\begin{figure}[tbph]
\begin{center}
\includegraphics[scale=0.35]{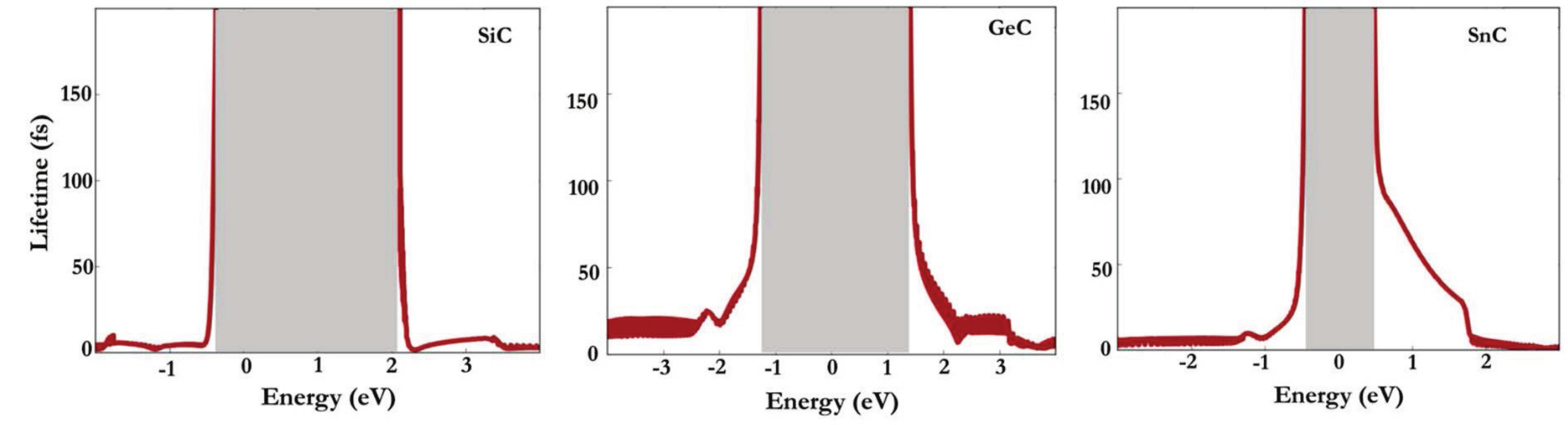}
\end{center}
\caption{{\protect \small Hot electron relaxation time corresponding to SiC,
GeC and SnC.}}
\label{x0004}
\end{figure}

Hot carrier relaxation time, that is inversely proportional to the imaginary
part of the self energy, shows in Fig.(\ref{x0004}) a high value near the
band edges. Moreover, electronic energy regions with many bands results in a
faster electron-phonon relaxation time of hot electrons by phonon emission.
This behaviour is due to the large number of possible transitions. The hot
electron lifetime plot corresponding to the SiC structure indicates that hot
carrier relaxation near the band edge is characterized by relaxation time
around 90 fs while faster relaxation time of 10-20 fs are found at energy
more than 75 meV away from the band edge. For GeC, the analysis of Fig.(\ref%
{x004}-d) reveals hot electron lifetime around 100 fs near the Fermi level
at room temperature. For an energy ranging from 200 meV up to the Fermi
level, the relaxation time is much shorter at less than 50 fs. Finally
around the edge of the band of SnC, the gap is characterized by a relaxation
time limiting the thermalization of the rate of hot electrons in the
interval ranging from 100 fs to 120 fs. At 100 meV away, the hot electron
relaxation time ranges in the interval [10,90] fs . In the three
carbon-based hybrids, the relaxation time observed near the band edges are
slower than the 400 fs found for graphene \cite{grappp}.

\subsection{Linewidth and scattering rate phonon dependency}

\begin{figure}[tbph]
\begin{center}
\includegraphics[scale=0.3]{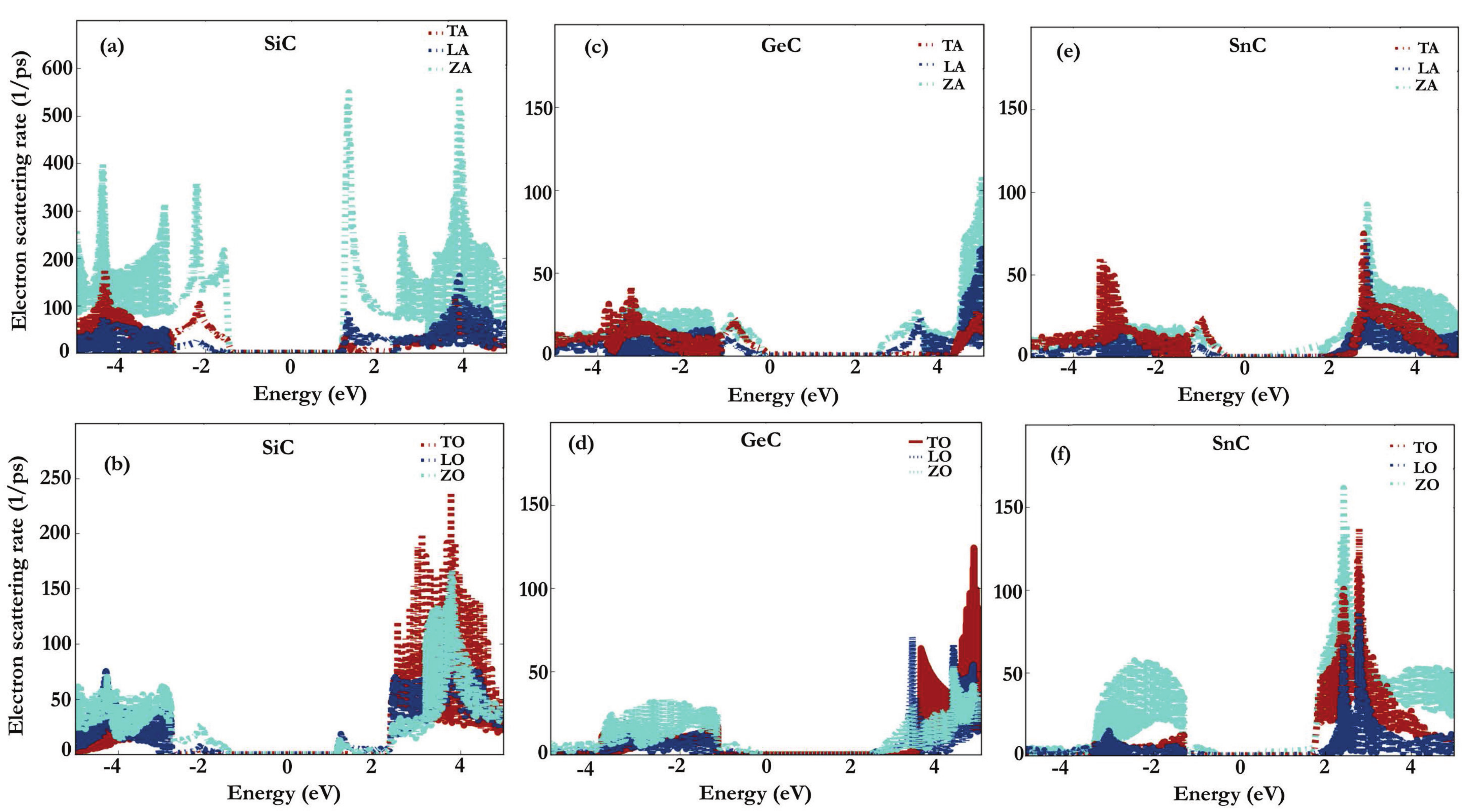}
\end{center}
\caption{{\protect \small Electron scattering rate associated with acoustic
(in the first line) and optical (in the second line) modes respectively for
SiC, GeC and SnC monolayers. }}
\label{x005}
\end{figure}

In general, the electron scattering rate arises from the contribution of
acoustic and optical phonons together. Fig.(\ref{x005}) provides a
quantitative analysis of the contributions from the individual phonon modes.
In SiC hybrid, the scattering rate is mainly dominated by the out of plane
acoustic mode $(ZA)$ over all energy levels. More precisely, $ZA$ mode is
followed by the transversal acoustic mode ($TA)$ mode at VBM, while $ZA$ is
followed by the longitudinal acoustic mode ($LA)$ mode at CBM. For higher
energy levels, $ZA$ is followed by the optical transversal mode $(TO)$. When
the Si atom is replaced by the much heavier Ge atom, the optical modes
contribution is getting much important. As shown in Figs.(\ref{x005}-c,-d),
the electron scattering is mostly due to the presence of $TO$ followed by%
\textbf{\ }$ZA$\textbf{\ }modes for higher electron energy ranges in GeC. At
the band edges of GeC, the CBM arise essentially from both $ZA$ and $ZO$
mode while the VBM is dominated by $ZA$ and $TA$. Unlike SiC, the major
contribution arises from the optical modes in SnC. Indeed, both the VBM and
the CBM are \ mainly dominated by $ZO$ mode followed by $TA$ mode. For
higher energy in the conduction band, $ZO$ is followed by $ZA$.

\begin{figure}[tbph]
\begin{center}
\includegraphics[scale=0.35]{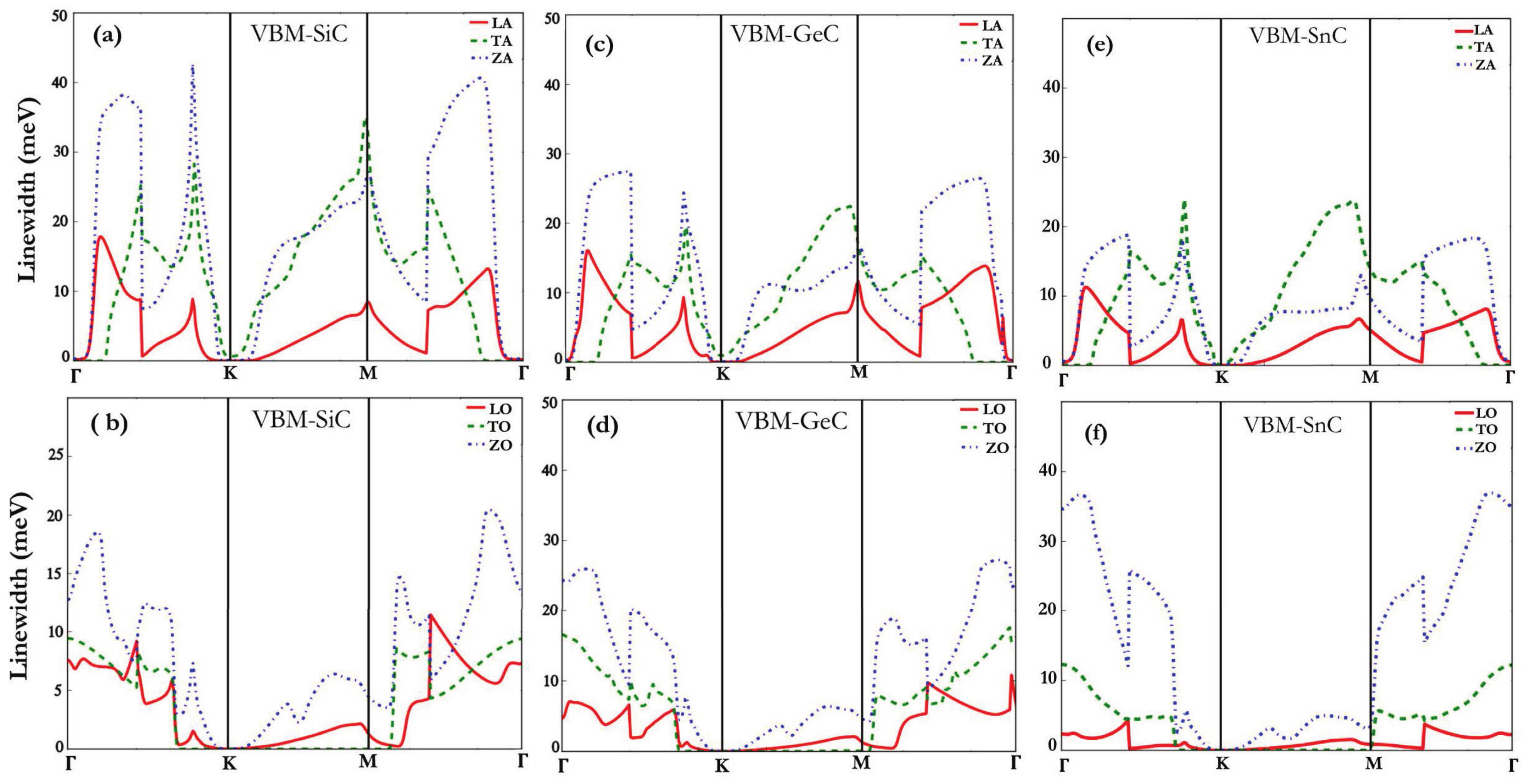}
\end{center}
\caption{Electron linewidth variation in terms of electron wavevector
\textbf{k} projected on the six phonon modes for SiC, GeC and SnC, where the
initial electron state is at the VBM. The contribution for each phonon mode
is displayed for the same electron energy for (a,c,e) acoustic phonon
branches LA (red continuous), TA (green dashed) and ZA (blue dashed-dotted);
and (b,d,f) the optical phonon branches LO (red continuous), TO (green
dashed) and ZO (blue dashed-dotted). On the top figures correspond to the
acoustic contribution and the bottom are the optical contribution. $\Gamma $%
, $K$ and $M$ describe the high symmetry point of electron wavevector. }
\label{x001}
\end{figure}

To shed more light on the \textbf{$k$}-dependence of the imaginary part of
the self energy in the three binary compounds, Figs.(\ref{x001}) and (\ref%
{x002}) display the linewidth variation over the high symmetry line of BZ
and electron state being at the $CBM$ and $VBM$. In these figures, the
linewidth is projected over acoustic and optical phonon modes to (i) show
their different contribution\textbf{\ }depending on the high symmetry
directions related to real space and to\textbf{\ }(ii) deduce information
regarding the linewidth and which phonon plays crucial role in this
direction.

\subsubsection{VBM}

For electron state at the $VBM$, the examination of acoustic modes
contributions plotted in Fig.(\ref{x001}) shows that the $LA$ mode is lower
compared to $TA$ and $ZA$ in all configurations. Over the $\Gamma -K$ and $%
M-\Gamma $ directions, the highest contribution in the linewidth is
attributed to the $ZA$ mode for SiC and also for GeC but with smaller
ampleness. This result matches very well with their counterparts silicene
and germanene \cite{ficheti}. In SnC, Fig.(\ref{x001}-e) reveals that the
linewidth arise mainly from both $ZA$ and $TA$ modes with a $ZA$ magnitude
very small compared to GeC and SiC. At $K$-point, all acoustic modes have no
contribution in the electron linewidth due to the weak interaction occurring
because of the lack of states at the band edges as shown in Fig.(\ref{x001}%
). \newline
For optical phonon, the curves plotted in Figs.(\ref{x001}-b,-d and -f) show
that in SiC, the linewidth is dominated by $ZO$ mode all over the high
symmetry line, followed by $LO$ and $TO$. In both GeC and SnC, the main
contribution arises also from the out of plane mode $ZO$.

\begin{figure}[tbph]
\begin{center}
\includegraphics[scale=0.35]{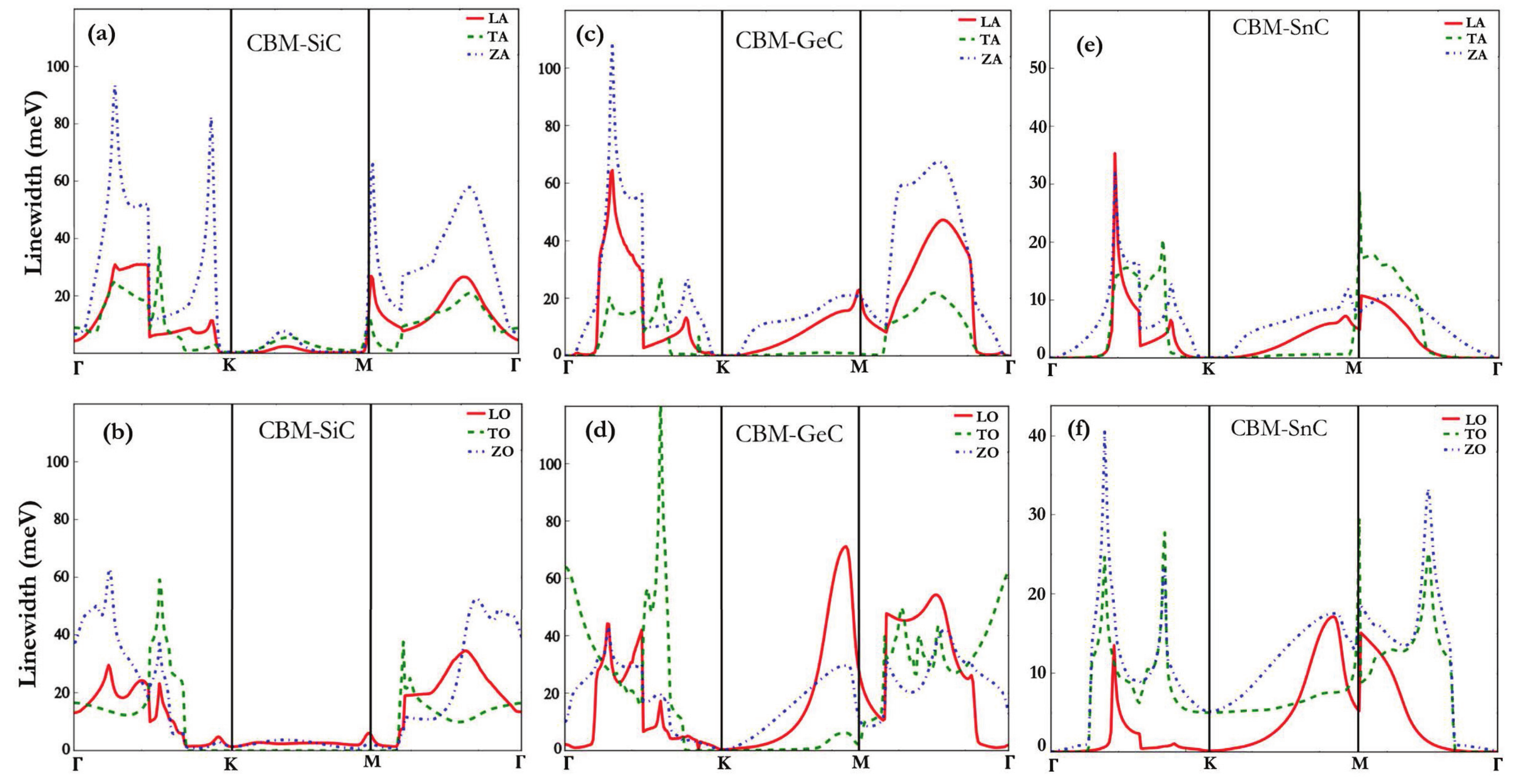}
\end{center}
\caption{Electron linewidth as a function of electron wavevector \textbf{k}
projected on the six phonon modes for SiC, GeC and SnC from left to the
right, where the initial electron state is at the CBM. The contribution for
each phonon mode is displayed for the same initial electron energy for
(a,c,e) acoustic phonon branches LA (red continuous), TA (green dashed) and
ZA (blue dashed-dotted); and (b,d,f) the optical phonon branches LO (red
continuous), TO (green dashed) and ZO (blue dashed-dotted). Top figures
correspond to the acoustic contribution while the bottom to the optical
contribution. $\Gamma $, $K$ and $M$ are the high symmetry points of
electron wavevector.}
\label{x002}
\end{figure}

\subsubsection{CBM}

For electron state at the CBM, Fig.(\ref{x002}) plots the electron linewidth
projected over phonon modes. In SiC, the largest contribution comes from $ZA$
mode over the high symmetry line, however the linewidth is shared into three
parts over the high symmetry line as displayed for GeC. Indeed, the
linewidth is dominated by the $TO$ and $ZA$\textbf{\ }in the $\Gamma -K$
region and only by $ZA$ in the interval $M-\Gamma $. While, the major
contribution is attributed to the $LO$ mode along $K-M$. In SnC, the
electron linewidth is dominated by $ZO$ mode for all the BZ high symmetry
line. It is worth noting that in SiC and GeC, the linewidth is zero for all
phonon modes at $K$-point. However, for both $TO$ and $ZO$ mode, the
linewidth is not zero in SnC. This difference is due to the indirect gap of
SnC as the band edge is located at $K$-point for VBM and at $\Gamma $-point
for CBM.

Finally, by evaluating the linewidth over the typical edge of the Brillouin
zone in SiC, GeC and SnC, one can deduce that there exists a large variation
in the magnitude of the electron-phonon coupling along the different
crystallographic directions.

\subsection{Conclusion}

Using ab-initio calculations, we reported a theoretical description of the
electron-phonon coupling effects on the gap, linewidth and scattering rate
in two-dimensional SiC, GeC and SnC materials. In addition to the study of
the thermal sensitivity of the linewidth, we investigated its k-resolved
behavior. This type of study is important for transport phenomena where
phonons play a crucial role and therefore has an impact on the engineering
concepts of devices based on nanostructures. While the band gap in SiC and
GeC showed a normal monotonic dependence to temperature, its behavior was
rather anomalous in the SnC semiconductor. It has been found that the
out-of-plane acoustic transverse mode $ZA$ efficiently couples to the
electronic states at the SiC band edge, while the linewidth mainly arises
from the $ZA$ and $ZO$ modes in GeC. In SnC, the optical mode, namely $ZO$,
was predominant. The hot carrier thermalization losses in SiC, GeC and SnC
under solar lighting were also calculated. It was found that, under sunlight
illumination, the hot carriers thermalize within 90 fs, 100 fs and 120 fs
for SiC, GeC and SnC respectively. This work highlights the microscopic
origin of the thermalization of hot electrons, which is difficult to deduce
from experiments and opens the way toward the study of materials for
renewable energies by means of ab intio calculations of hot electrons.

\begin{acknowledgement}
L. B. Drissi and N. B-J. Kanga would like to acknowledge "Acad\'{e}mie
Hassan II des Sciences et Techniques-Morocco" for financial support. L. B.
Drissi, F. Djeffal and S. Haddad acknowledge the Extended African Network
for Advanced 2D Materials funded by ICTP, Trieste, Italy for networking. L.
B. Drissi, S. Lounis and F. Djeffal thank also the Arab German Young Academy
of Sciences and Humanities (AGYA) sponsored by the Federal Ministry of
Education and Research (BMBF) for the membership scheme. S. L. acknowledges
support from the European Research Council (ERC) under the European Union%
's Horizon 2020 research and innovation programme (ERC-consolidator
Grant No. 681405 DYNASORE).
\end{acknowledgement}

\end{document}